\documentclass[]{aastex631}

\newcommand{\hd}{HD$\,$169142}
\newcommand{\bertrang}{\cite{2018MNRAS.474.5105B}}
\newcommand{\bertrangp}{\citep{2018MNRAS.474.5105B}}

\shorttitle{On moving shadows and pressure bumps in HD$\,$169142}
\shortauthors{Bertrang et al.}
\graphicspath{{./}{figures/}}

\begin{document}

\title{On moving shadows and pressure bumps in \hd \footnote{Based on observations made with ESO Telescopes at the Paranal Observatory under program 0101.C-0595(A).}}

\email{bertrang@mpia.de}

\author{Gesa H.-M. Bertrang}
\affiliation{Max Planck Institute for Astronomy \\
K\"onigstuhl 17 \\ 
69117 Heidelberg, Germany}

\author{Mario Flock}
\affiliation{Max Planck Institute for Astronomy \\
K\"onigstuhl 17 \\ 
69117 Heidelberg, Germany}

\author{Miriam Keppler}
\affiliation{Max Planck Institute for Astronomy \\
K\"onigstuhl 17 \\ 
69117 Heidelberg, Germany}

\author{Trifon Trifonov}
\affiliation{Max Planck Institute for Astronomy \\
K\"onigstuhl 17 \\ 
69117 Heidelberg, Germany}

\author{Anna B. T. Penzlin}
\affiliation{Institut f\"ur Astronomie \& Astrophysik, Universit\"at T\"ubingen \\ Morgenstelle 10 \\ 
D-72076 T\"ubingen, Germany}

\author{Henning Avenhaus}
\affiliation{Max Planck Institute for Astronomy \\
K\"onigstuhl 17 \\ 
69117 Heidelberg, Germany}

\author{Thomas Henning}
\affiliation{Max Planck Institute for Astronomy \\
K\"onigstuhl 17 \\ 
69117 Heidelberg, Germany}

\author{Matias Montesinos}
\affiliation{Departamento de Ciencias B\'asicas, Universidad Vi\~na del Mar,\\
Vi\~na del Mar, Chile}
\affiliation{ N\'ucleo Milenio de Formaci\'on Planetaria (NPF), Chile}

\begin{abstract}

The search for young planets had its first breakthrough with the detection of the accreting planet PDS70b. In this study, we aim to broaden our understanding towards the formation of multi-planet systems such as HR8799 or the Solar System.
Our previous study on HD169142, one of the closest Herbig stars, points towards a shadow-casting protoplanetary candidate. Here, we present follow-up observations to test our previously proposed hypothesis. We set our new data into context with previous observations to follow structural changes in the disk over the course of 6 years. We find spatially resolved systematic changes in the position of the previously described surface brightness dip in the inner ring. We further find changes in the brightness structure in azimuthal direction along the ring. And finally, a comparison of our SPHERE data with recent ALMA observations reveals a wavelength dependent radial profile of the bright ring.
The time-scale on which the changes in the ring's surface brightness occur suggest that they are caused by a shadow cast by a 1--10M$_J$ planet surrounded by dust, an orbit comparable to those of the giant planets in our own Solar System. Additionally, we find the first indications for temperature-induced instabilities in the ring. And finally, we trace a pressure maxima, for the first time spatially resolved, with a width of 4.5$\,$au. The density distribution of the ring at mm~wavelengths around the pressure maxima could further indicate effects from snow lines or even the dynamics and feedback of the larger grains.

\end{abstract}



\section{Introduction} \label{sec:intro}

\begin{figure}
\centering
\plotone{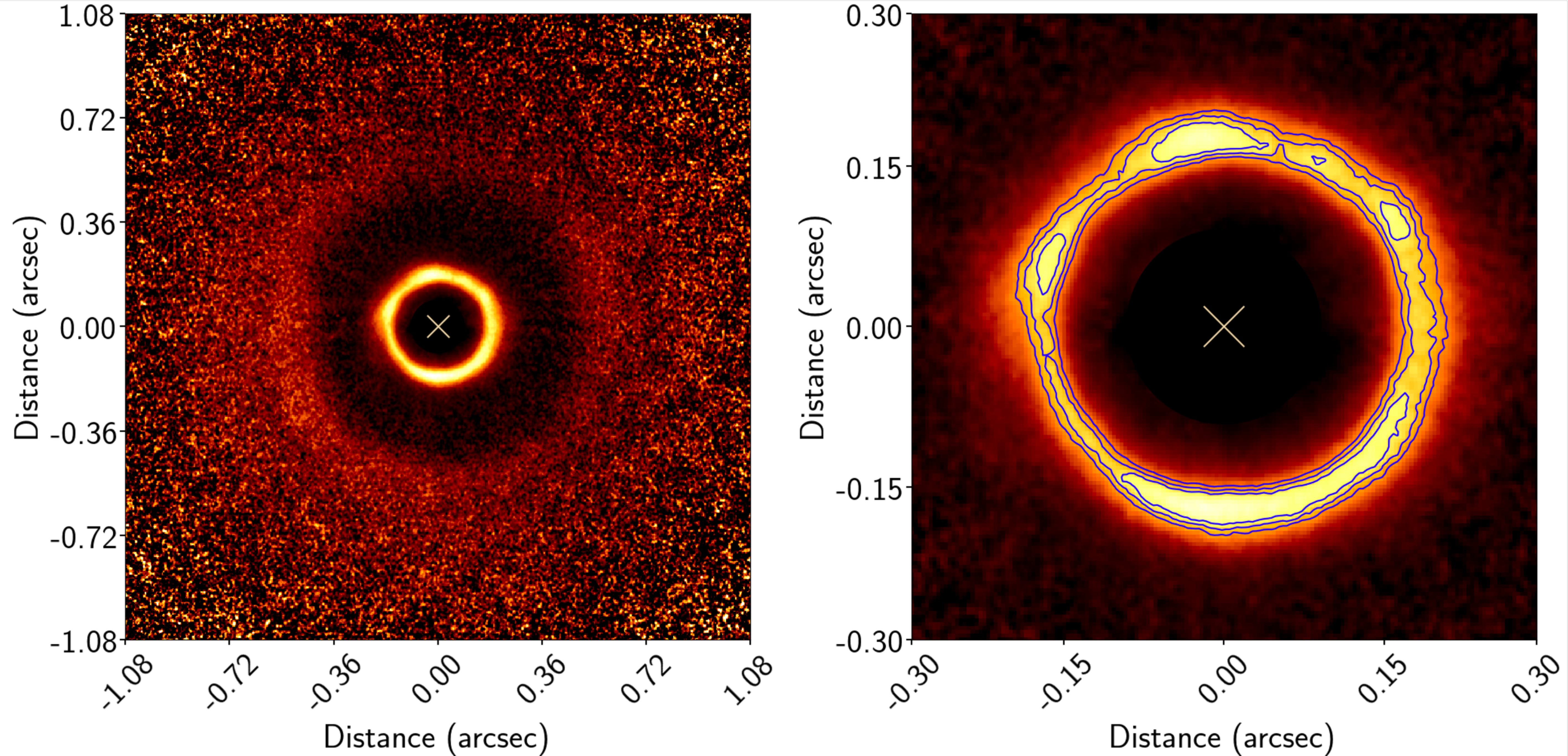}
\epsscale{0.95}
\plottwo{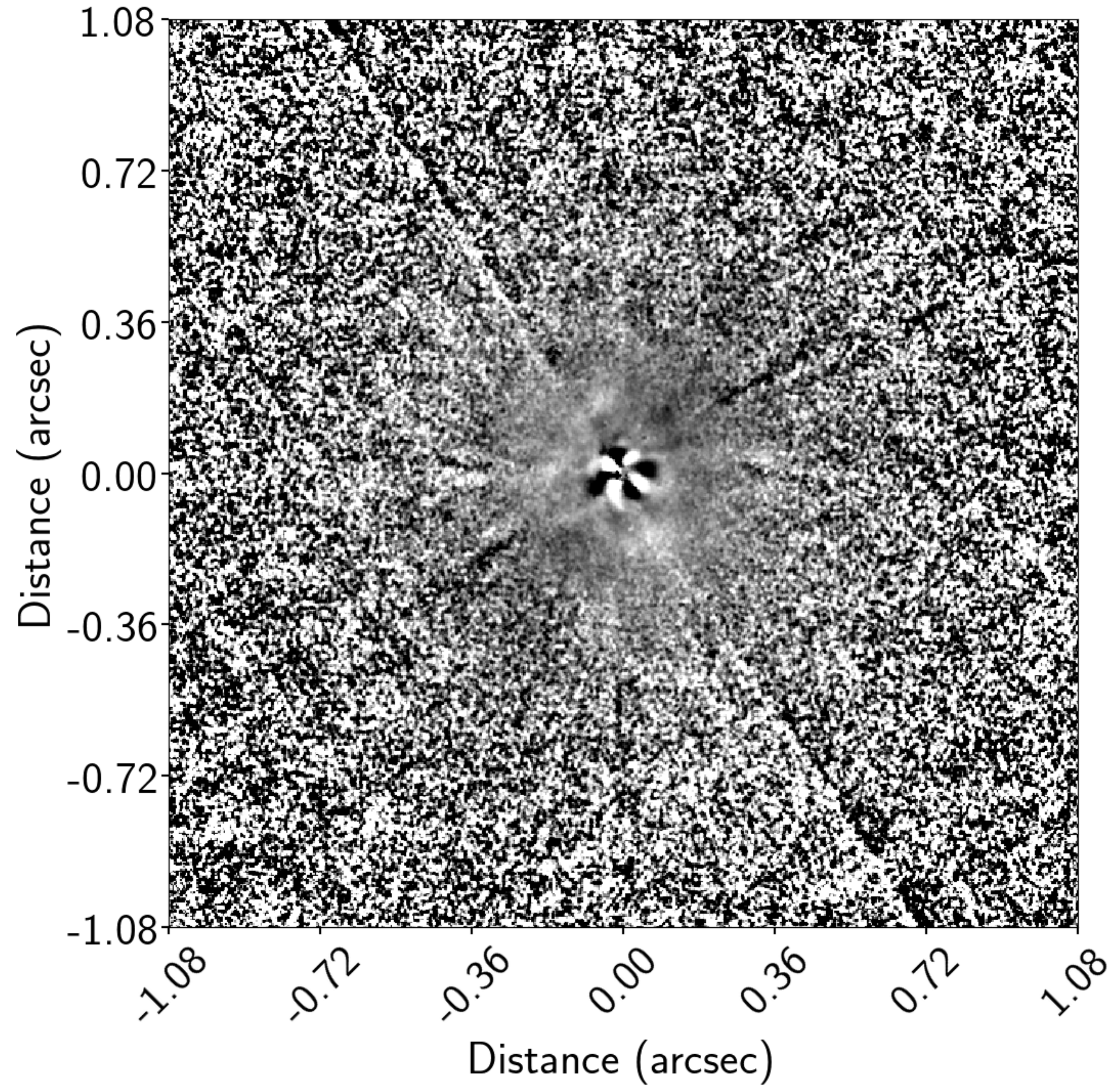}{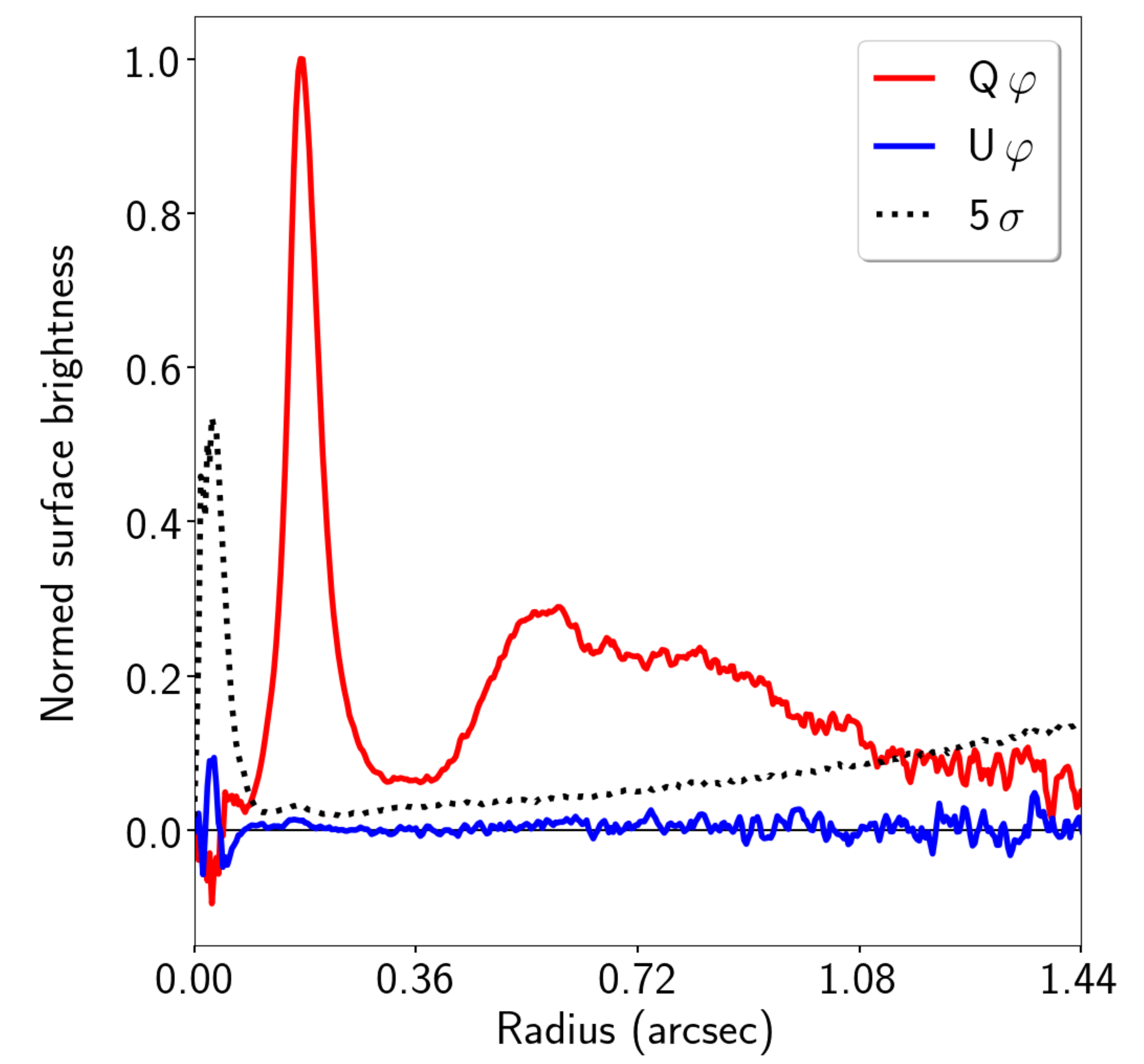}
\caption{ZIMPOL/SPHERE PDI observations of HD$\,$169142 in VBB. {\it Upper Left:} The final $Q_\varphi$ map. {\it Upper Right:} A zoom onto the bright ring in $Q_\varphi$. Contour lines indicate 30 per cent, 35 per cent, and 40 per cent levels of the peak flux. The inner $0.1\,$arcsec of this observation are dominated by the butterfly-noise pattern (see lower left panel), typical for PDI observations, and thus,  blanked out. {\it Lower Left:} The final $U_\varphi$ map. All maps are displayed in $r^2$ scaling, North is up and East is left. {\it Lower Right:} Azimuthally averaged radial surface brightness profile of $Q_\varphi$ (red solid line) and $U_\varphi$  (blue solid line), as well as $5\sigma$~level (black dashed line).}
\label{pic:obs1}
\end{figure}

The Herbig Ae star HD~169142, with a mass of $ {1.65\pm 0.05}\,$M$_{\odot}$ \citep{2018arXiv180209280C} at a distance of $113.5\,$pc  \citep[the parallax is $8.77 \pm 0.06\,$mas; ][]{2016A&A...595A...1G, 2018A&A...616A...1G, 2018AJ....156...58B}, is surrounded by a complex structured protoplanetary disk. The disk structures that are traced since the arrival of extreme AO~instruments such as SPHERE \citep[Spectro-Polarimetric High-contrast Exoplanet REsearch;][]{2019A&A...631A.155B} at the Very Large Telescope (VLT) or the Atacama Large (sub-)Mm Array (ALMA) allow us to probe for the physics of planet formation in previously unreached detail \citep[e.g.,][]{2018ApJ...863...44A, 2018ApJ...869L..41A, 2020A&A...633A..82G, 2020arXiv200105007A}.

Still rarely observed but of all the greater interest are time-variable structures. Usually dominated by their orbital time-scale, the difficulty to detect such features lies in the observational limitations of today's instrumentation close to the star. However, if features on an orbit close to the star affect the more outer disk, such as in the case of shadow-casting, then time-variable structures become observable already now \citep[][]{2017ApJ...849..143S, 2018ApJ...857L...9R}. In this study, we present a new case of time-variable disk features in the disk around \hd. This well-studied disk is composed out of several gaps and disk segments, most prominent: the bright inner ring around $\sim0.18\,$arcsec ($\sim20\,$au) as well as an outer disk starting at $0.47\,$arcsec ($55\,$au) from the star \citep[e.g.,][]{2018MNRAS.474.5105B, 2019arXiv190707277M}. 

 Several studies, performed at various wavelengths, conclude on three proto-planetary candidates within this disk, located at  $\sim13\,$au,  $\sim34\,$au, and  $\sim46\,$au \citep[][{\it this work}]{2017A&A...600A..72F, 2018MNRAS.474.5105B,  2019A&A...623A.140G}. There are two more candidates discussed in the literature, one of them likely being a disk feature \citep{2014ApJ...792L..22B, 2014ApJ...792L..23R, 2018MNRAS.473.1774L, 2019A&A...623A.140G, 2019MNRAS.486.3721B}. The other proposed candidate is motivated by recent ALMA observations, spatially resolving the outer disk into three rings \citep[][]{2019AJ....158...15P}. However, we would like to point out two cavities of this proposed scenario: (i) the brightness ratio of the three rings is not consistent between the observation and the model and (ii) the modelled gap ({\it dent}) along the planetary orbit, which the planet carves into its direct surrounding, is neither on the same ring nor of the same size as in the observation by \cite[][]{2019AJ....158...15P}. Future studies may test the proposed scenario of a mini-Neptune in the outer disk in more detail, however, this is beyond the scope of this study.
 
 In \bertrang, we compared ZIMPOL data of \hd~ with earlier VLT/NACO observations \citep{2013ApJ...766L...2Q} and found indications of a systematically moving surface brightness dip in the bright ring. We determined the angular velocity of the dip to $\sim10^{\circ}/$yr. Based on this, we proposed that this dip may be a shadow cast by a Jupiter-mass/brown dwarf object, surrounded by a dust disk or envelope, at a distance of only $6\,$au to the star \bertrangp. To test our hypothesis, we repeated our SPHERE/ZIMPOL observation three years later and present here both, the new data along with a new model for this disk.
 
In Section~\ref{sec:obs}, we describe the observations and data reduction. The results and analysis of the observations are described in Section~\ref{sec:res}. We discuss our results and conclude in Section~\ref{sec:dis} and Section~\ref{sec:con}.

\section{Observations and Data Reduction}\label{sec:obs}

\begin{deluxetable}{c c c c c c c}
\tablecaption{Observation summary.\label{obstab}}
\tablewidth{0pt}
\tablehead{
\colhead{Night} & \colhead{Filter~$1$} & \colhead{Filter~$2$} & \colhead{Read-out} & \colhead{Polarization} & \colhead{DIT} & \colhead{time}\\
\nocolhead{} & \nocolhead{} & \nocolhead{} & \colhead{mode} & \colhead{mode} & \colhead{(s)} & \colhead{(s)}
}
\startdata
July~$9, 2015$ & VBB *  & VBB *  & fastPol & P$2$ & 4.38 & $3154\,$\\
\enddata
\tablecomments{ * The R and I band filters were interchanged during the observation, but the data was reduced together.}
\end{deluxetable}

The data were taken during the night of July 15, 2018, at the ESO Very Large Telescope on Cerro Paranal, Atacama desert, Chile (see Tab.~\ref{obstab}). The instrument used was the Zurich Imaging Polarimeter (ZIMPOL) sub-instrument of SPHERE. The data were taken in the P2 mode (for more details, consult the SPHERE user manual). The filter used was the VBB/RI filter (very broad band encompassing both the R and I band, central wavelength: $0.73\,\mu$m, see SPHERE manual). The observations were taken in FastPol mode to allow for optimal contrast. 

This observation is a follow-up on our previous findings \citep{2018MNRAS.474.5105B} with the aim to probe for structural changes during the time-window of $3\,$years.
In order to be as comparable as possible, we used the same observation block (OB; slightly reduced integration time to account for service mode observation rules) from our previous program. Further, we applied the same steps for data reduction and computation of the local Stokes vectors  $Q_{\varphi}$ and $U_{\varphi}$ as described in \cite{2018MNRAS.474.5105B}.

In Section~\ref{sec:res}, we present archival ALMA data that have been originally published in  \cite{2019AJ....158...15P}. Here, we use the calibrated data as provided in the ALMA archive.

\begin{figure}
\centering
\plottwo{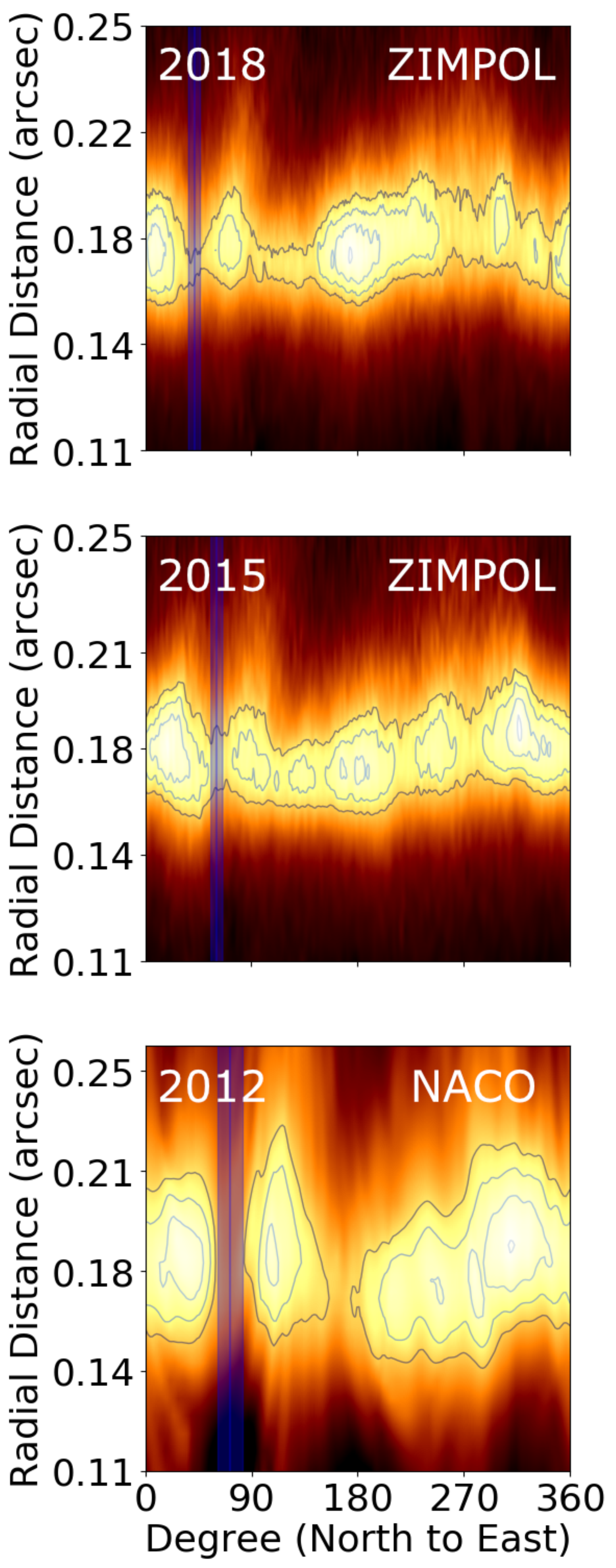}{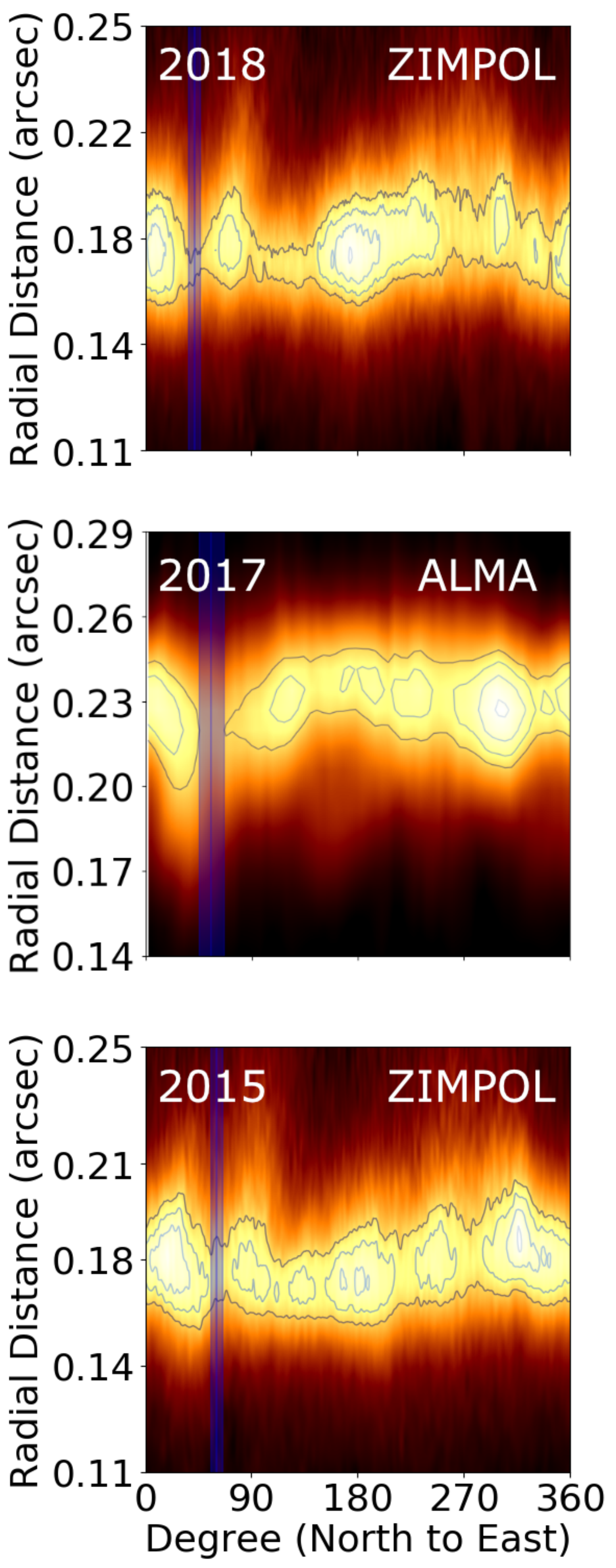}
\caption{{\sl Right: }The bright ring in polar coodinates observed in 2012 with VLT/NACO \citep[{\it bottom;}][]{2013ApJ...766L...2Q}, in 2015 \citep[{\it mid;}][]{2018MNRAS.474.5105B} and 2018 with VLT/SPHERE-ZIMPOL ({\it top; this work}). Contour lines indicate \{68, 78, 88, 98\} per cent levels of the peak flux in the ring. The blue shaded region marks the brightness dip moving with the Keplerian velocity of a shadow-casting Jupiter-mass candidate at a radial distance of $\sim0.1\,$arcsec from the star.\\
{\sl Left: } The bright ring in polar coodinates  observed with VLT/SPHERE-ZIMPOL in 2015 and 2018 ({\it top, bottom;} as in Fig.~\ref{pic:polarplot}), compared with ALMA observations at $1.3\,$mm from 2017  ({\it mid}). Contour lines indicate $\{68, 78, 88, 98\}$ per cent levels of the peak flux in the ring. The blue shaded region marks the brightness dip as in Fig.~\ref{pic:polarplot}. As expected from a shadow cast onto the ring, the temperature and therefore, the brightness in the ALMA map drops below the shadowed area.}
\label{pic:polarplot}
\end{figure}

\section{Results and analysis}\label{sec:res}

\subsection{Disk morphology and brightness asymmetries}

The $Q_{\varphi}$ and $U_{\varphi}$ images are shown in Figures~\ref{pic:obs1}. The data are not flux calibrated, thus, we are limited to an analysis of the relative surface brightness distribution. 

 We trace the disk down to $\sim0.12\,$arcsec ($\sim14\,$au) to the star; closer to the center, the data is dominated by the butterfly-noise pattern which is typical for PDI observations. The azimuthally averaged radial surface brightness is shown in Figure~\ref{pic:obs1}, bottom. The overall structure remains stable in comparison to previous observations at comparable spatial resolution ($0.0230 \times 0.0233$ arcsec$^2$; $2.69 \times 2.72\,$au) and wavelength \citep[cf.][]{2018MNRAS.474.5105B}: A bright ring peaks around $0.18\,$arcsec ($\sim20\,$au), followed by a gap which stretches from $\sim0.25-0.38\,$arcsec ($\sim28-43\,$au). The outer disk, detected up to $1.08\,$arcsec ($\sim123\,$au) in our observation, appears continuous, intermitted by a brightness dip around $0.72\,$arcsec ($\sim82\,$au). For a detailed analysis of the disk's overall architecture, see \bertrang.
 
In the following sub-sections, we present comprehensive evidence based on multiple, independent factors which strongly support our hypothesis of a new, shadow-casting candidate.

\subsection{Shadow cast onto the ring}

\begin{figure}
\centering
\epsscale{0.5}
\plotone{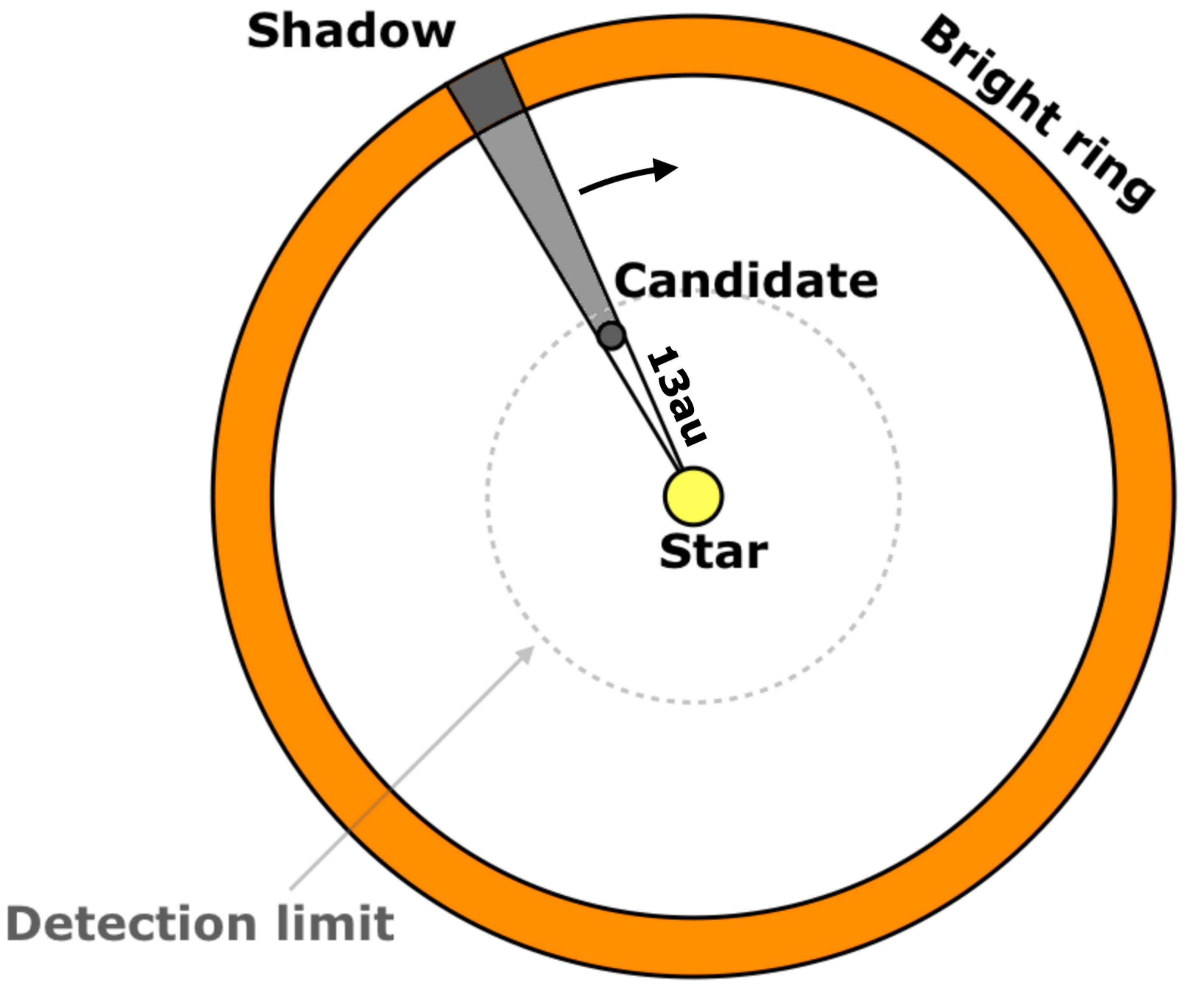}
\caption{Scheme of the shadow cast by the proto-planetary candidate onto the bright ring. In-depth multi-wavelength modelling found this disk to be particularly flat \bertrangp, and therefore, allows us to conclude on the CPD parameters by simple geometric optics. The dotted ring indicates the error dominated area in which our observations are not sensitive to.}\label{pic:sketch}
\end{figure}

In Figure~\ref{pic:polarplot}, three maps of $Q_{\varphi}$ in polar coordinates, observed over the course of 6~years, are shown. The maps are zoomed onto the bright ring. Since the observations in 2012 were obtained with VLT/NACO in H~band, this map is less well-resolved compared to the maps observed with VLT/SPHERE in the optical. However, a comparison between all three observations shows that the main features are conserved.

The bright ring is strongly sub-structured, its peaks and dips have a contrast ratio of more than $30\%$. One particular dip caught our attention when we compared the 2012 and 2015 observations. \cite{2013ApJ...766L...2Q} report a dip in the surface brightness profile at $\sim80^{\circ}$ which appeared to have moved in three years by $\sim30^{\circ}$ \bertrangp. Our new observations, obtained with another offset of three years reveal a dip in the surface brightness which is consistent with the one observed three and six years earlier. This dip has a width of $\sim 0.027\,$arcsec and is spatially resolved in both observations performed with VLT/SPHERE. The high spatial resolution of the new observation allows us to narrow down the center position of the dip to $42\pm5^{\circ}$. Compared to the position of the dip observed in 2015 ($61\pm5^{\circ}$), and assuming Keplerian velocity as well as a circular orbit, the dip moves with $\sim6^{\circ}/$yr too fast for the radial position of the ring where it is detected. However, this surface brightness dip is moving with the velocity of an Keplerian object on an orbit at  
$\sim0.12\pm 0.09\,$arcsec$\,(\sim13.4\pm 10.36\,$au). Only the very upper limit is in the range of the Keplerian velocity for the radial position at which we find the dip, future observations will allow us to narrow down the error on this value (see Appendix~\ref{app:errors}). Concluding, this dip is consistent with a Jupiter-mass protoplanetary candidate surrounded by dust casting a shadow onto the bright ring (see Fig.~\ref{pic:sketch}) as suggested by \bertrang\footnote{A detailed theoretical study on shadow-casting planets is beyond the scope of this study.}

As shown in \cite{2017ApJ...837..132V}, the scattering surface of a disk can be as low as $1-2$~pressure scale-heights above the mid-plane. Our estimate on the Hill sphere of this candidate (Sect.~\ref{sec:hill}) gives a pressure scale-height of the shadow-casting candidate of $0.6-1.3$. Hence, a shadow cast by this candidate is a likely explanation of the observed brightness dip.

In addition to the shadow, we find more structural changes along the ring with the same time-scale as the movement of the shadow. A potential explanation of this observed feature is the temperature-induced instability \citep{2018ApJ...860...27I}. Based on our previous modelling of this disk \citep{2018MNRAS.474.5105B, 2019arXiv190707277M}, we can make an estimate of the thermal relaxation time-scale \citep{2015ApJ...811...17L, 2017ApJ...850..131F}. The rough estimate derived by us,  $\tau_{relax}\sim1\,$yr, has to be set in context to the dynamical time-scale. The thermal time-scale is the time it takes for the disk to heat or cool, and the dynamical time-scale is the time for the disk to reach hydrostatic equilibrium. If both time-scales are equal, the temperature-induced instability reaches its maximum. We know from 3D~dynamical simulations \citep{2011ApJ...735..122F} that the dynamical time-scale at the scattering surface of disks is of comparable order and therefore, the temperature-induced instability is indeed a potential explanation of the observed features. This finding might motivate future work on this effect which is beyond the scope of this study.

\subsection{Impact on the mid-plane}

\begin{figure}
\centering
\epsscale{0.5}
\plotone{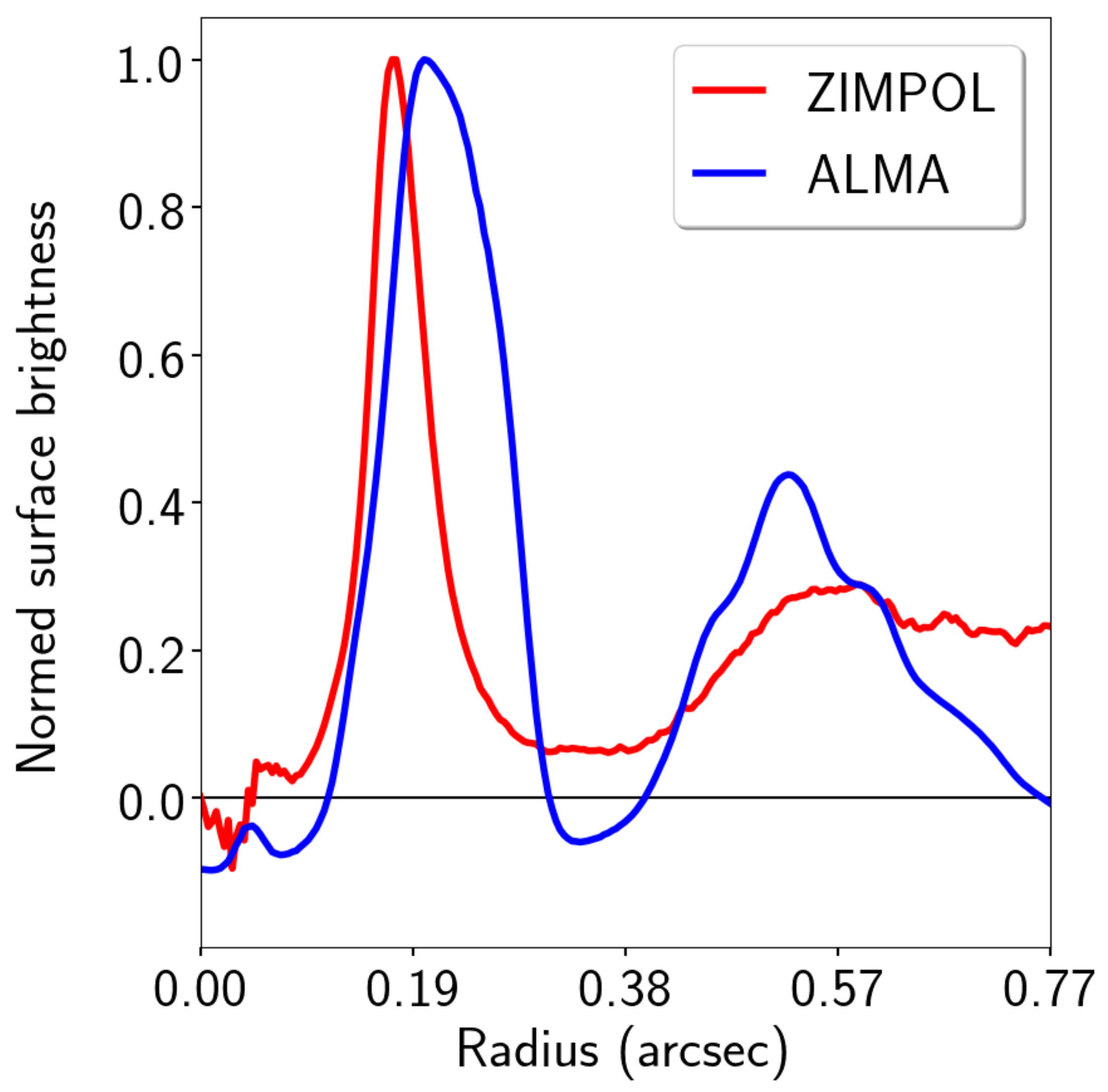}
\caption{Azimuthally averaged radial surface brightness profile as observed with SPHERE/ZIMPOL in 2018 (red;  $Q_\varphi$ as in Fig.~\ref{pic:obs1}, \ref{pic:obs1}) and ALMA at Band 6 (blue). The shift in radial position of the bright ring, indicating a pressure bump is clearly visible (see Section~\ref{bump}). In the outer disk, the three rings described in \cite{2019AJ....158...15P} cover the outer disk and its substructures as seen in the optical at the same radial positions.}\label{pic:radprof}
\end{figure}

Our VLT/SPHERE-ZIMPOL observations remain the observations of HD$\,$169142 with the highest spatial resolution up-to-date.  However, recently published ALMA observations obtained in 2017 at $1.3\,$mm \citep{2019AJ....158...15P} trace this disk at a resolution only 2-times more coarse than our optical resolution. This enables us, for the first time, to directly compare the data at these two wavelengths. For this purpose, we show the archival ALMA data chronologically ordered together with the ZIMPOL data in polar coordinates in Figure~\ref{pic:polarplot} as well as a direct comparison of the azimuthally averaged radial brightness profiles in Figure~\ref{pic:radprof} (see also Section~\ref{bump}).

The bright ring shows clear sub-structures of contrasts of more than $30\%$ also in the ALMA map. If our interpretation of the brightness dip in the optical is correct and it is indeed a shadow cast by an object on an inner radius, then the shadow is expected to cool the region it is cast on. This drop in temperature will be detected as a dip in the brightness distribution probed with ALMA. And indeed, this is what is detected. In 2017, the dip in the mid-plane is located at $55.5\pm11^{\circ}$.

\subsection{Surface density rise and pressure bump}\label{bump}

When comparing the bright ring as observed in the optical and the mm (see Figure~\ref{pic:polarplot} and Figure~\ref{pic:radprof}), two things become obvious: (i) going to longer wavelengths, the ring moves outwards. We detect the radial center of the ring in the optical positioned around $0.18\,$arcsec while it appears in the mm regime centred around $0.22\,$arcsec. This is a shift of $4.5\,$au and marks the position between small and large grains; small grains scatter light the strongest at the increase of the gas surface density and large grains emit the strongest at their peak concentration which is located at the pressure maximum. Such a configuration is expected for a disk profile with a steep surface density rise with larger grains trapped in a pressure bump \citep[e.g., ][]{2016A&A...590A..17R, 2017ApJ...850...52P, 2019A&A...624A...7V}. Further we note that both brightness profiles show an interesting feature:  (ii) The ring has a sharp and a smooth edge. In the optical, the sharp edge is at the inner rim of the ring while, at $1.3\,$mm, it is the ring's outer rim which is sharp and the smooth edge points towards the ring's position at optical wavelengths.

While the findings of the optical surface brightness profile are expected, the radial profile of the mm emission points to a situation where the grains a mixed not equally on both sides of the pressure trap. This is very interesting and could indicate effects from snow lines or even the dynamics and feedback of the larger grains.  

\subsection{CPD size and planet-mass from shadow-width}\label{sec:hill}

The ability to cast a shadow relies on a certain geometrical size and optical depth of the casting object (see Figure~\ref{pic:sketch}). Hydrodynamical simulations suggest that the matter accreted by a planet is, rather than in an envelope, distributed in a circumplanetary disk \citep[CPD; e.g.,][]{2016MNRAS.460.2853S}.  By applying simple geometrical optics, we derive a CPD~radius of $\sim1.1\,$au. Various simulations point to the fact that the CPD will have a radius between $0.3-1\,$r$_{\rm hill}$  \citep[e.g.,][]{2009MNRAS.397..657A, 2009MNRAS.393...49A, 2011MNRAS.413.1447M, 2012A&A...548A.116R, 2016MNRAS.460.2853S}. Using this as base, we derive a mass of $\sim1-30\,$M$_{\rm J}$ for the shadow-casting planetary candidate. However, a mass of $\gg10\,$M$_{\rm J}$ is rather unlikely as the system could become easily unstable (see Sec.~\ref{sec:dis}). Future observation will be able to test for this scenario.

\begin{deluxetable}{ccccc}
\tablecaption{Parameter space of the orbital stability analysis.\label{tab}}
\tablewidth{0pt}
\tablehead{
\colhead{Semi-major axis} & \colhead{Position angle} & \colhead{Eccentricity} & \colhead{Mass} & \colhead{References} \\
\colhead{(au)} & \colhead{(deg)} & \colhead{Number} & \colhead{(M$_{\rm J}$)} & \nocolhead{References}
}
\startdata
$12\pm2$ & $55.5\pm4$ & $0.05\pm0.05$ & $7\pm3$ & 1\\
$35.5\pm0.5$ & $34.9\pm1.4$ & $0.05\pm0.05$ & $2\pm1$ & 2, 3 \\
$45\pm5$ & $224\pm4$ & $0.05\pm0.05$ & $2\pm1$ & 2, 3\\
\enddata
\tablerefs{ (1) This work (2) Gratton~et~al.~(2019)  (3) Bertrang~et~al.~(2018)}
\end{deluxetable}

\section{Discussion}\label{sec:dis}

This disk is object of many hours of observations across the wavelength spectrum. However, this is the first time data was gathered over the course of $3$~epochs, at multiple wavelengths and similar spatial resolution, to be analysed in a temporal context\footnote{This study is qualitatively different from the discussion in \cite{2018MNRAS.473.1774L} and \cite{2019A&A...623A.140G}.}.  The detection of the features described in Section~\ref{sec:res}, the moving shadow, the temperature-induced turbulences, and the characterisation of the pressure bump have only been possible due to this observational effort over time and across the spectrum.

\begin{figure}
\centering
\epsscale{0.5}
\plotone{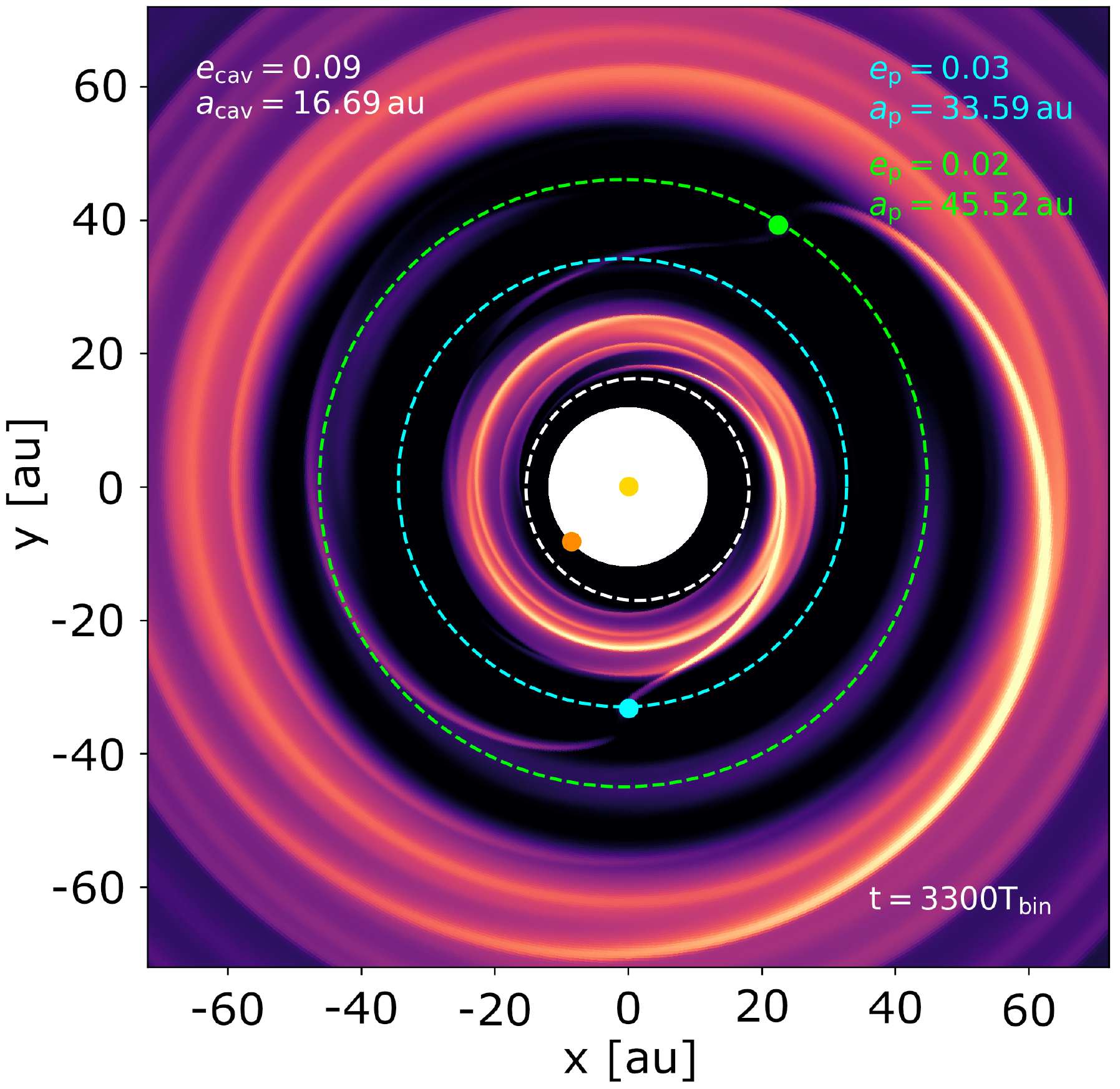}
\caption{Our new hydrodynamical simulations, including migration, of 3 planets ({\it orange}: $10\,$M$_{\rm J}\,\rm{at}\sim12\,$au; {\it blue}: $1\,$M$_{\rm J}\,\rm{at}\,\sim34\,$au; {\it green}: $10\,$M$_{\rm J}\,\rm{at}\,\sim46\,$au) orbiting around the central star ({\it yellow}). The size of the inner cavity ({\it white circle}) as well as the width of the bright ring and the gap are equivalent to the those found in \hd. These simulations show that the architecture of the disk around \hd~can be explained with the shadow-casting candidate and two Jupiter-mass planets in the outer gap.}
\label{pic:hydro}
\end{figure}

In general, a single companion with a mass larger than \mbox{$4\cdot10^{-3}\,$times} the mass of the central star is able open an eccentric disk cavity \citep[e.g.,][]{2006A&A...447..369K, 2018A&A...616A..47T}. Given the evidence we find for the shadow-casting candidate, we test (a) whether this candidate alone is able to open the inner, slightly eccentric disk cavity and (b) whether the overall architecture of the disk around \hd~ can be explained by a migrating three-planet scenario. For this,  we perform a locally isothermal 2D~hydrodynamical simulation using the GPU version \citep{2017A&A...604A.102T} of the PLUTO code \citep{2009MSAIS..13...67M}. For the setup, we follow the procedure as described in \cite{2018A&A...616A..47T} and additionally, we set $H/R = 0.05$, the viscous parameter $\alpha = 10^{-3}$ and the disk mass $\mathrm{M_{disc}} = 0.01 \mathrm{M_{star}}$. We assume the inner planet has an orbit with zero eccentricity. We exclude the inner $12\,$au from the simulation grid for performance reasons. However, disk feedback on N-body objects is included for the full system. All bodies are non-accreting. The result of this hydrodynamical simulation is shown in Figure~~\ref{pic:hydro}.

We begin these simulations with only the star and the inner planet at 12 AU, the shadow-casting candidate, for $50'000$ orbits. After 5000 orbits, we find a convergent behaviour of the inner cavity with an eccentricity oscillating between $e = 0.05 - 0.11$ and a cavity radius of $20 - 22\,$au. These parameters fit very well to the observed shape of the cavity \citep[see also][]{2018MNRAS.474.5105B}. A first attempt to open a gap with only one outer planet failed to explain the width of the observed gap. Therefore, we insert two planets of $1$M$_{\rm J}$ each at $35\,$au and $45\,$au after $50'000$ orbits into the disk and let them clear their orbit without feedback from the disk for $2000\,$orbits. Those radial positions are motivated by  previous studies \citep{2018MNRAS.474.5105B, 2019A&A...623A.140G}. Then, we enable disk feedback on the planets and let them migrate for $4000\,$orbits. The planets are still slowly migrating with a rate of $0.15\,$au over 100 orbits for the middle planet.

Our simulations show that a single massive planet at $12\,$au can indeed cause the slightly eccentric inner cavity as it is observed in \hd. Further, two massive planets are necessary to open the observed gap starting at about 25~au. 

The stability of such a multiple giant proto-planet system is somewhat questionable. Therefore, we test the orbital stability of the three-planet system with N-body simulations using the Wisdom-Holman algorithm \citep[][]{Wisdom1991}. We generate $100\,000\,$normally distributed samples of orbital parameters and planetary masses based on the results of previous studies (see Table~\ref{tab}). For each sample, we run an N-body simulation of $10\,$Myr, which is the estimated age of this system \citep[][]{2017ApJ...850...52P}. The time step of the Wisdom-Holman integrations was set to $90\,$days, which we find to be fully sufficient for an accurate orbital simulation. We consider the system unstable if at any given time one of the planets reaches an extremely high eccentricity of e$\,>0.99$ (i.e., a prerequisite of highly-unstable crossing orbits), or the planetary distance to the star exceeds three times the initial semi-major axis of the outermost planet (i.e., a planetary ejection).

\begin{figure}
\centering
\epsscale{1}
\plottwo{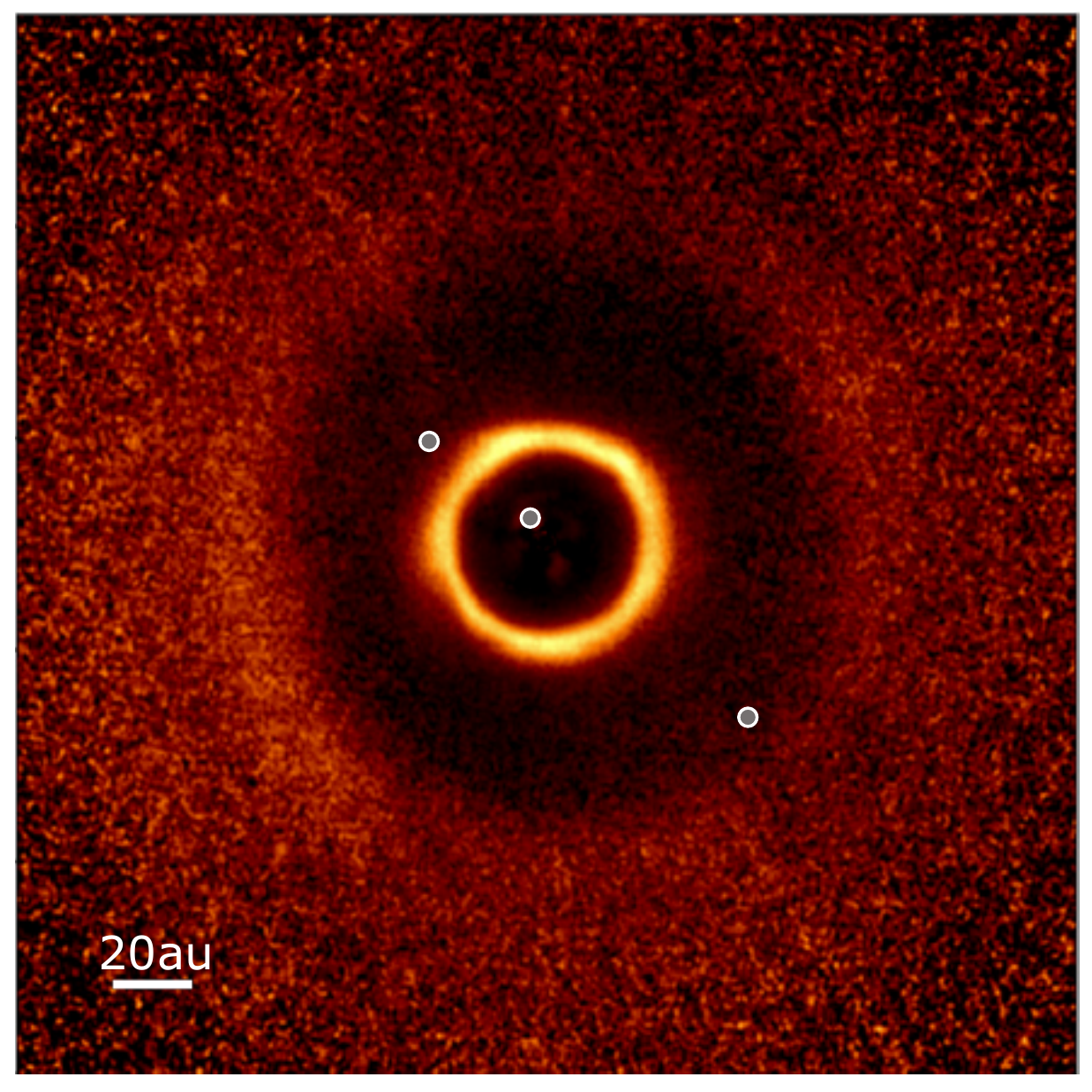}{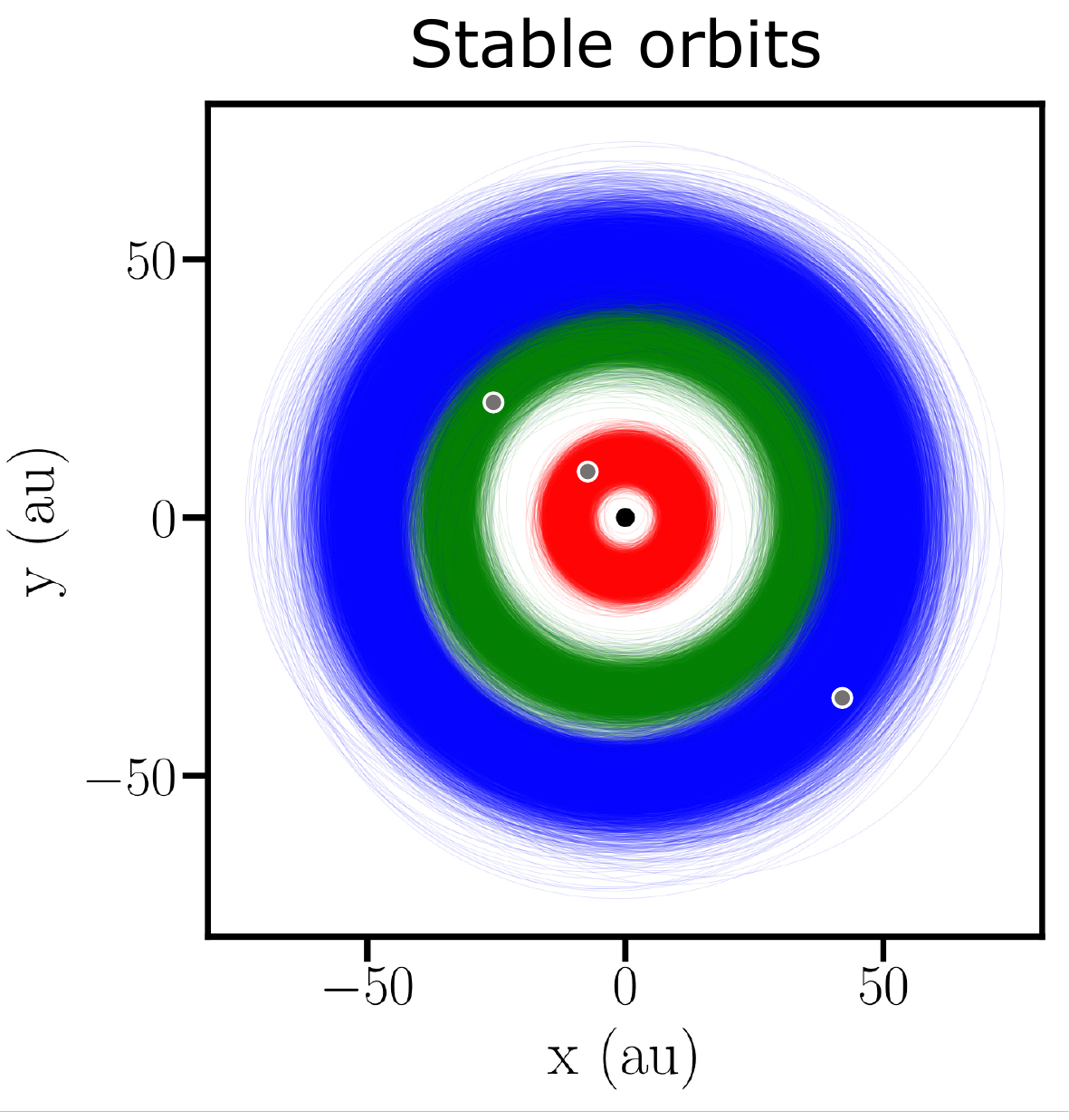}
\caption{{\it Left:} The ring and gaps observed in this disk are most consistent with a multiple massive planets (white-rimmed grey circles), therefore, HD$\,$169142 might be a young version of HR$\,$8799 or a young cousin of our own Solar System. {\it Right:} Our stability analysis of such a multi-planet system results in a large number of stable orbits (from the inner to the outer planets (white circles) in the top panel: blue, green, red) that fit into the architecture of this disk (black circle: central star).}
\label{pic:orbits}
\end{figure}

The resulting stable orbits are in very good agreement with our findings (Fig.~\ref{pic:orbits}). This allows us to put effective constraints on the planetary masses, semi-major axes, and eccentricities. The proposed architecture of this system relies on planetary masses of $\lesssim10$~M$_{\rm J}$ and small eccentricities to be stable. The semi-major axis of the outer most protoplanetary candidate could not be below $45\,$au. Future observation will be able to validate this hypothesis.

We would like to emphasise that the protoplanetary disk around \hd~is only marginally flared \bertrangp. Therefore, the circumplanetary disk candidate casts a shadow  onto the bright ring efficiently also if it is oriented co-planar with the protoplanetary disk.

Summarizing, we present the proposed architecture of this protoplanetary system in Figure~\ref{pic:orbits}. The new shadow-casting candidate at only $12\,$au, an orbit comparable to those of the giant planets in our own Solar System, is added as well. More observational evidence is needed, however, to confirm whether HD$\,$169142 is indeed harbour to a young multiple massive planet system, and therefore, young version of a multi-planet systems such as HR$\,$8799 or our own Solar System.

%

\section{Conclusion}\label{sec:con}
We present new VLT/SPHERE-ZIMPOL data on HD$\,$169142. This data set is the most recent in a cycle of three observations, covering $6$~years. We further compare the optical data with new ALMA observations, and test for the stability of the proposed multi-planet system hiding in this disk. Our main results are the following:

\begin{itemize}
\item The surface brightness dip in the bright ring described in \cite{2013ApJ...766L...2Q} and \bertrang~ appears also in this new observation at the expected position. Therefore, a shadow-casting $\sim1-10\,$M$_{\rm J}$ planet, surrounded by dust, at $0.12\,$arcsec ($13\,$au), an orbit comparable to those of the giant planets in our own Solar System, is consistent with our findings. 

\item We find a brightness dip in the ALMA observation at $1.3\,$mm which is consistent with the position of the brightness dip (in the optical) and thus, with a shadowed and therefore cooled region in the mid-plane of the disk.

\item A comparison between the optical and the mm data reveals that the large grains in the ring are shifted by $4.5\,$au further out  (in comparison with the small grains) which clearly points to a local pressure bump caused by a rise in surface density.  

\item Further structural changes in the ring are observed on the same time-scale as the potential shadow is moving, however, with much less systematics in shape. This is consistent with temperature-induced instabilities and the first observational indication for this instability.

\item Our new hydrodynamical simulation show that the observed structure of \hd~is consistent with a system of 3 migrating massive protoplanets at $\sim13\,$au ($10\,$M$_{\rm J}$), $\sim34\,$au ($1\,$M$_{\rm J}$), and $\sim46\,$au ($1\,$M$_{\rm J}$).

\item Our stability analysis of the proposed multiple massive planet system matches the proposed positions of the protoplanetary candidates and we show the stable orbits in Figure~\ref{pic:orbits}. Effective constraints from our N-body test on the planetary masses, semi-major axes, and eccentricities are: planetary masses of $\lesssim10$~M$_{\rm J}$ and small eccentricities are necessary to stabilize this system and the semi-major axis of the outer most protoplanetary candidate could not be below $45\,$au. Future observation will be able to validate this hypothesis.

\end{itemize}

By now, there is plenty of indirect evidence for a young multiple massive planet system orbiting HD$\,$169142, however, direct evidence is still pending. Therefore, even though the data set is already vast, it is necessary to gather additional data especially aiming at the innermost region around the star. At this point, \hd~is the most  convincing candidate for a young version of multi-planet systems such as HR~8799 or our own Solar System.

\begin{acknowledgements}
      GHMB and MF acknowledge financial support from the \emph{European Research Council (ERC)} under the \emph{European Union’s Horizon 2020 research and innovation programme} (grant agreement No. 757957). MM acknowledges financial support from the Chinese Academy of Sciences (CASSACA) Postdoctoral Fellowship in Santiago, Chile, and from the Iniciativa Cient\'ifica Milenio via the N\'ucleo Milenio de Formaci\'on Planetaria (NPF). TH acknowledges support from the European Research Council under the Horizon 2020 Framework Program via the ERC Advanced Grant Origins~83~24~28. This research has made use of the {\sc Exo-Striker} toolbox \citep{Trifonov2019es}, which uses a custom version of the {\sc Swift} N-body library \citep{Levison2013}. This work has made use of data from the European Space Agency (ESA) mission
{\it Gaia} (\url{https://www.cosmos.esa.int/gaia}), processed by the {\it Gaia}
Data Processing and Analysis Consortium (DPAC,
\url{https://www.cosmos.esa.int/web/gaia/dpac/consortium}). Funding for the DPAC
has been provided by national institutions, in particular the institutions
participating in the {\it Gaia} Multilateral Agreement.
\end{acknowledgements}

\bibliographystyle{aasjournal}
\bibliography{biblio}

\begin{appendix} 
\section{Keplerian velocity of the bright ring and  the surface brightness dip}\label{app:errors}
Aiming to understand the underlying nature of the observed surface brightness dip in the bright ring of this disk, we examine the Keplerian velocity of an object at the radial position of the dip in the bright ring and compare it to the azimuthal velocity of the observed dip. 

We evaluate both the azimuthal velocity of the surface brightness dip and the Keplerian velocity at a reference position,  $r_{\rm{ref}}$. The corresponding physical quantities and their errors are given in Table~\ref{errtab}. We apply the rules of propagation of uncertainty to compute the error for any quantity derived from the basic measurements.

As reference position, we choose the radial center of the ring, $r_{\rm{ref}} = 0.175\,$arcsec which, given the parallax, corresponds  to
\begin{equation}
	r_{\rm{ref}} = \left(19.94\pm 0.71\right)\,\rm{au.}
	\label{eq:rref}
\end{equation}

The azimuthal velocity of the surface brightness dip is
\begin{equation}
	v_{\rm az} = \omega \cdot r_{\rm ref} = \left(10.45\pm 4.26 \right)\,\rm{km/s}, 
	\label{eq:vaz}
\end{equation}
where  $\omega =\varphi/t$ is the angular velocity.

In order to compare the azimuthal velocity, $v_{\rm az}$, of the observed surface brightness dip to the Keplerian velocity, $v_{\rm{Kepler}}$, at the reference position $r_{\rm{ref}}$, we compute
\begin{equation}
	v_{\rm{Kepler}} = \sqrt{\frac{G\cdot M}{r_{\rm{ref}}}} = \left(8.57\pm0.02 \right)\,\rm{km/s}, 
	\label{eq:kep}
\end{equation}
where $G$ is the gravitational constant and $M$ the stellar mass.

Under the assumption of Keplerian motion, we can derive the orbital distance of an object, $r_{\rm obj}$, with the azimuthal velocity $v_{\rm az}$ from Equation~(\ref{eq:vaz}):
\begin{equation}
	r_{\rm obj} = \frac{G \cdot M}{v_{\rm az}^2} = \left(13.4\pm 10.39 \right)\,\rm{au} \equiv \left(0.12\pm 0.09 \right)\,\rm{arcsec} 
	\label{eq:orbobj}
\end{equation}

Comparing the results from the calculation of velocities (Eq.~(\ref{eq:vaz}), (\ref{eq:kep})) or the radial positions (Eq.~(\ref{eq:rref}),(\ref{eq:orbobj})), it is clear that, while Keplerian motion of the dip consistent with the ring cannot be ruled out, the Keplerian velocity at the reference point in the ring is only at the very outer limit of the error on the azimuthal velocity of the observed surface brightness dip. The movement of the dip is much more consistent with a Keplerian motion on an orbit much closer to the star, as shown in Equation~(\ref{eq:orbobj}). To narrow down the errors on the observables, however, more data is needed.

\begin{deluxetable}{c c c c}[b]
\tablecaption{Physical quantities and their errors.\label{errtab}}
\tablewidth{0pt}
\tablehead{
\colhead{Physical quantity} & \colhead{Variable} & \colhead{Value} & \colhead{Reference}
}
\startdata
Stellar mass & $M$  & $(1.65\pm0.05)\,$M$_{\odot}$ & 1 \\
Parallax & $p$  & $(8.77 \pm 0.06)\,$mas & 2, 3, 4 \\
Time between the observations * & $t$  & $(3\pm0.019)\,$yr & 5 \\
Change of azimuthal position * & $\varphi$  &  $(19\pm 7.07)^{\circ}$ & 5 \\
\enddata
\tablecomments{ *measured for the surface brightness dip as observed in 2015 and 2018.}
\tablerefs{ (1) Gratton~et~al.~(2019) (2) GAIA Collaboration~(2016) (3) GAIA Collaboration~(2018) (4) Bailer-Jones~et~al.~(2018) (5) This work.}
\end{deluxetable}

\end{appendix}

\end{document}